\documentclass{optica-article}

\journal{opticajournal} 

\articletype{Research Article}

\usepackage{lineno}

\begin{document}

\title{Noise Reversal by Entropy Quantum Computing}

\author{Yu-Ping Huang \authormark{1,$\dagger$} Yongxiang Hu \authormark{2,*}}

\address{\authormark{1}Quantum Computing Inc., 5 Marineview Plaza, Hoboken, USA, 07030\\
\authormark{2} NASA Langley Research Center, Hampton, Virginia 23681, USA\\}

\email{\authormark{$\dagger$}yhuang@quantumcomputinginc.com\\
\authormark{*}yongxiang.hu-1@nasa.gov} 


\begin{abstract*} 
Signal to noise ratio is key to any measurement. Recent progress in semi/super-conductor technology have pushed the signal detection sensitivity to the ultimate quantum level, but the noise issue remains largely untouched and, in many cases, becomes even more severe because of the high sensitivity. In this paper, we explore a hardware-based approach to noise removal using entropy quantum computing. Distinct to any existing de-noising approach, it observes and reproduces the quantum statistical properties of noise in an optical system to emulate and thereby reverse the noise from data. We show how it can recover 1D and 2D image data mixed with much stronger noise. 
\end{abstract*}

\section{Introduction and Philosophy}
When Richard Feynman first proposed the concept of quantum computing, his vision was about building a controllable system to ``make a simulation of nature'', for problems intractable for classical computers \cite{Feyman1982}. Since then, a large variety of algorithms have been developed to solve NP hard problems, most requiring fault-tolerant, gate-based quantum computers with substantial variable numbers and circuit depth (see \cite{QC2011,QC2017,horowitz2019quantum} and references therein). As such, while exciting progress has been made over the past decade in demonstrating qubit coherence, integration, logic operations, and error correction, there has yet to be a demonstration where gated-based quantum computing is superior in solving problems of practical relevance \cite{sood2024archives}. 

In this paper, we revisit the great thoughts behind the conception and exertion of quantum computing, and propose a convincing use case where quantum technology does create practical values. To claim any quantum value, the total cost and operation overhead associated with the deployment of a quantum device must be outweighed by the critical advantages it brings to an essential mission, when comparing with the best classical counterparts. In this assessment, a golden criteria is the net quantum advantage, defined as the performance enhancement minus the price paid, including the device parameters of size, weight, power, cost (SWaP-C) and the required supporting resources. In other other words, the system performance shall be wholistically benchmarked against the classical state of the art, not just on a select of indices. For a net-positive advantage, its performance must surpass the classical counterparts under similar SWaP-C constraints. 

We explore a potential use case of quantum computing with net-positive advantages for remote sensing over extended distances. Rather than focusing on some interesting math problems, our starting point is an imminent problem facing aeroborne and spaceborne missions: small signal and high noise \cite{CALIPSO09,ARMS18}. For example, space light detection and ranging (LiDAR) systems does not work well in daytime because the photon-level returning signals can be mixed with strong reflected sunlight \cite{Daytime15}. One way to address this issue is to enhance the signal by increasing the probe laser power. This, however, will incur high costs in launching and significant complications in operation, by virtual of the exceeding needs for electric power and cooling. 

This motivates our proposal: instead of enhancing the signal, why not decreasing the noise? There are two primary avenues to it. The first is to filter out the noise as much as possible, by using tight filters to reject noise not in the quantum modes of the returning signal \cite{hlawatsch1994time}. For LiDAR, the quantum modes refer to time bins, spectral bands, orbital angular momenta, polarizations, and any hybrid combination of them. Yet there is an inherent trade-off between the filtering efficiency and the insertion loss of the signal as a result \cite{slepian1961prolate}. Recently, we demonstrated that quantum parametric mode sorting can achieve the optimum trade-off allowed by Heisenberg uncertainty \cite{shahverdi2017quantum}. Yet the implementation is not trivial and may require advanced photonic integrated circuits to reduce the SWaP-C parameters for space missions \cite{rehain2020noise,kumar2018mode}. 

Here we seek an approach alternative and complimentary to the noise filtering. The fact is, regardless of how good filters are, there are always noise that are in exactly the same mode of the signal and thus cannot be rejected. The detected results are thus always a mixture of the signal and noise, which are indistinguishable. The million-dollar question is then: could we still use other differential properties to distinguish the signal and noise?

The answer is yes. Depending on the system parameters and application scenarios, the quantum statistical properites of the signal and noise could be adequately different. This is what behind quantum entanglement sensing, where the signal is entangled with a party (often named as the idler in the quantum information community) (see \cite{degen2017quantum} and references therein). At the receiver, a joint measurement designed around this entanglement favors the detection of the signal over the noise that is not entangled \cite{gatti2004ghost,shih2012physics,shapiro2020quantum}. The problem is, however, that entanglement is challenging to produce, and the rate is often prohibitively low. As such, the detection advantage has remained on a same-quanta basis, and does not translate to performance enhancement for practical uses. A good example is quantum illumination, where quantum entanglement can increase the likelihood of detecting a true signal photon over a background photon by a factor equal to the number of modes the signal spans \cite{lloyd2008enhanced}. This is indeed an advantage over classical sensing by itself. The catch, unfortunately, is that only a single photon is used as the probe at one time, so that it takes a exceedingly long time to build up appreciable statistics, even if the returning loss is low (opposite to most LiDAR applications). Instead, classical LiDAR uses bright pulses containing millions to billions of photons. The measurement signal to noise per pulse is easily much higher than quantum illumination. This fundamental problem, coupled with system complexity and unfavorable SWaP-C parameters, makes quantum entanglement sensing questionable. Note that this deficiency may be overcome by using high photon-number correlated states over a large number of modes; however it has not been experimental demonstrated \cite{huang2021method}. 

In this paper, we propose to address the noise challenge differently. Instead of employing entanglement, we use the conventional LiDAR but exploit the differential statistical properties of the signal and noise. To this end, we assume the following:
\begin{enumerate}
    \item There must be certain levels of spatiotemporal correlation among the returning signals, as long as the targets have some spatial continuity. In contrast, noise usually does not have such correlation, due to the inherent randomness of its generation and detection.  
    \item The noise obeys Poissonian statistics, by virtue of being "shot noise", with its variance equal to the mean \cite{henry1996quantum}. 
    \item The total amount of noise is known, which can be measured either during the off-period or in between pulses, if the signal is in pulses.
\end{enumerate}
These assumptions apply to most active sensing techniques, LiDAR and Radar included. Our goal is to devise a technique just based on them, without any further assumption or information about the targets. This distinguishes it from any machine learning methods, where lots of trainings are needed and misinformation might be created. 

Our approach is based on noise retrieval and reversal. That's, given a set of measured data (which could be 1D, 2D, or 3D), let's find the corresponding noise for each data point such that:
\begin{itemize}
    \item the noise obeys Poissonian distribution;
    \item all noise in a group adds up to a known total;
    \item the noise-subtracted data have high spatiotemporal correlation.
\end{itemize} 
This is a NP-hard optimization problem. Even if the Poissonian statistics is not account for, there are an exponential number of possible noise distributions, making calculations extremely difficult, if not impossible, to be performed in a Turing machine . 

This is where quantum computing comes to rescue. One way is to turn this into a standard Hamiltonian and solve it in a gate-based or annealing-based system. However, such will require lots of qubits and considerable depth or connectivity for practical applications where the data points and noise quanta can each exceed thousands easily \cite{sood2024archives}. 

Instead, our solution is to build a specialty quantum machine to simulate the noise. This idea goes back to the embarking mission for quantum computing: to simulate the nature \cite{Feyman1982}! In particular, we proposal to build a quantum hardware system with many ``particles'' residing in many modes, where  
\begin{itemize}
    \item each particle resembles a noise quantum;
    \item each mode represents a data point;
    \item the particle number per mode obey the Poissonian statistics.
    \item all particles in certain groups add up to the known noise total in corresponding modes;
\end{itemize}
Then, the system would evolve to find the optimum distribution of the noise, such that after the noise-subtracted data gives the maximum spatial coherence. This is distinct to any algorithm-based method for denoising. Here, the noise is retrieved in a quantum emulator, and thereby reversed from the measurement data. That's why we coin this hardware solver as ``noise reversal''. 

In the following, we will describe this approach using a specific example of noise reversal for space LiDAR data, with in mind that the technique can be applied directly or after modification to many other applications. 

\section{Problem Setting}
Space LiDAR is a central tool of the Earth science observing technology \cite{national2007earth}.
It provides unambiguous vertical distributions of the ocean and atmospheric system, such as temperature, water vapor, wind speed, aerosol, cloud, and phytoplankton. While the scientific community relies on space LiDAR for unique scientific information of the weather and climate system, NASA faces significant challenges implementing these LiDAR missions. The need to filter out sunlight noise drives up the cost of such lidar missions by increasing the complexity of the lidar designs. In particular, in the low Earth orbit, the chance of a probing photon scattered by ocean/atmosphere and reaching the telescope of a space lidar is around $10^{-12}$. To achieve the desired signal-to-noise ratios (SNR's), space lidar missions require exceeding laser power to achieve its scientific objectives, which is very costly. 

Space LiDAR systems can usually provide high quality measurements at night. On the other hand, it is difficult for space lidars (such as CALIOP/CALIPSO) to detect aerosols during daytime due to sunlight noise \cite{thorsen2015calipso}. The CALIOP/CALIPSO satellite measurements initiated a novel and invaluable seventeen-year 3-D global aerosol and cloud climate record \cite{winker2009overview}. The reduced sensitivity to aerosols during daytime affects the quality of this climate record. Techniques to reduce sunlight noise in lidar images are needed to improve aerosol detection and reduce uncertainties in estimating aerosol radiative forcing and aerosol impact on air quality.

So far, spatial averaging is the most commonly used denoising method for both designing space lidar instruments and space-lidar operational data analysis. Yet, the advantage comes at the price of reduced spatial resolution of the science data. In most cases, high laser power and/or a large telescope are still required to achieve the desired SNRs for space lidar missions, which can significantly increase mission cost and take space lidars out of the cost caps of NASA’s Earth Venture (EV) and Earth System Explorer (ESE) missions. 

\section{Entropy Quantumm Computing}
Optimization problems involving discrete or continuous variables, some falling within the NP-hard or NP-complete class in the complexity theory, are prevalent in various domains of importance, such as operations, drug discovery, wireless communications, finance, integrated circuit design, compressed sensing, machine learning, and space exploration \cite{barahona1982computational}. Despite significant progress in algorithms and digital computer technology, even moderately sized NP-hard/NP-complete problems encountered in practical scenarios can prove highly challenging for modern digital computers \cite{garey1979computers}. To address this, one alternative method is stepping away from the traditional von-Neumann based approaches, to revisit analog computing \cite{rubio2019water}. This approach leverages the dynamic convergence of multiple competing phenomena towards an attractor, which serves as a representation of a function's extremum. 

This concept aligns with the notion of analog quantum computing, which maps a mathematical model into the Hamiltonian of a quantum system \cite{mukherjee2015multivariable}, and find the optimum solution by evolving the system into its ground state. One technique is use quantum annealing, which is a form of adiabatic computing \cite{das2008colloquium}. The drawback, however, is that such systems require cryogenic cooling and total isolation, while suffering limited qubits numbers and connectivity. Recently, entropy quantum computing (EQC) was proposed and demonstrated using quantum photonics, to offer efficiency, scalability, and SWaP-C qualifications for real-world problems \cite{nguyen2024entropy}. Distinct from competing techniques implemented in a closed, well-isolated quantum system, EQC is implemented in an open quantum system consisting of quantum optical signals coupled to an ‘environment’ of many degrees of freedom. It takes the challenges of such coupling with the environment and uses these interactions to relax the quantum signals to their collective ground states.


EQC uses “weak coupling” to the environment in a series of rapid measurements, causing the system to evolve into the least loss state via the quantum Zeno effect \cite{kwiat1995interaction,hosten2006counterfactual}. Once obtained, the solution state will remain in a decoherence-free subspace, thereby mitigating or avoiding the requirement for quantum error correction \cite{beige2000quantum}. An advantageous implementation of EQC is in quantum photonic systems, where among others, quantum Zeno Blockade can serve as a building block to construct an optimization solver or a gate machine \cite{sun2013photonic}. 


Based on EQC, Quantum Computing Inc. has developed its Dirac product series for discrete-variable optimization using an optical fiber loop and photon counting \cite{nguyen2024entropy}. In Dirac-3 machines, EQC starts with vacuum fluctuations subtending many quantum modes in the form of optical time bins. The quantum states in those modes are amplified and undergo a linear transformation. Then, they are passed through a loss medium with differential loss rates that depend on the collective quantum states. The output is then fed back to the amplifier to complete a loop. The loops are iterated till a stable state is reached, which represents a solution to the optimization problem with the minimum cost function. As such, EQC is efficient in finding the optimum solutions for binary, integer-number, and mixed integer problems. Among them, Dirac-3 can find the minimum cost function $H(V_i)$ for discrete variables $\{V_i\}$, with 
\begin{equation}
    H=\sum_{i} C_i V_i +\sum_{i,j} J_{ij} V_i V_j+\sum_{i,j,k} T_{ijk} V_i V_j V_k+\dotsc
    \label{eqH}
\end{equation}
where $C_i$, $J_{ij}$, $T_{ijk}$ are the linear energy (chemical potential), second-order nonlinear energy (two-body interaction), and third-order nonlinear energy (third-order interaction), respectively. $\{V_i\}$ are non-negative integers subject to a sum-constrain, i.e., $\sum_i V_i$ is a constant. 

In the machine, $V_i$ resembles the number of photons in a nanosecond time bin, and the cost function is translated to the photon loss for each time. Through loop iterations, the photons align themselves across the time bins to minimize the loss, and eventually settle at collective state with the minimum total loss. In this way, Dirac-3 naturally implements a discrete-variable solver for optimization problems. Not only it can go beyond the two-body interaction, but it also supports all-to-all connections. 

\section{Noise Reversal}
Here, we propose to realize an EQC-based emulator to retrieve the noise, by utilizing the one-to-one correspondence between the photons in the Dirac-3 machine and the noise. Considering a photon-counting LiDAR image with a total of $P$ pixels, with $M_i$ photons measured in the $i$-th pixel, and the total number of background photons is estimated to be $N$. In the emulator, 
\begin{itemize}
    \item there are effectively $N$ photons in the loop, each representing one background photon;
    \item there are $P$ time bins, each corresponding to a pixel of the image;
    \item by using coherent states in each time bin, the photon number distribution is naturally in Possionian distribution;
    \item the objective is to find the optimum photon numbers $\{N_i\}$ for the $i$-th time bin, such that the noise-subtracted data $\{M_i-N_i\}$ yield the highest spatial correlation.    
\end{itemize}
The above is clearly an NP hard problem, as the number of possible arrangement increases exponentially with $P$ and $N$. While it quickly becomes intractable using classical computers as $N,P$ reaches 100, the proposal quantum machine can efficiently retrieve the noise and relax to an optimum solution. 

There are multiple ways to define the cost function as a measure of the spatial correlation, depending on the correlation's degree and extend across pixels. For example, if the spatial correlation is localized, it can be captured well by a cost function with only the nearest neighbor interaction. If the correlation is of long range, the all-to-all connectivity may be warranted. Also, the correlation can be of different orders, to best reflect the target properties according to some a prior information. In this study, we will use the simplest cost function to capture the nearest neighbor correlation to the second order, with
\begin{equation}
    C=\sum_{i} \left(M_i-N_i-\frac{M_{i-1}-N_{i-1}+M_{i+1}-N_{i+1}}{2}\right)^2,
    \label{eqC}
\end{equation}
under the constraint of $\sum_i N_i=N$. 

To find the optimal $\{N_i\}$ in Dirac-3, we first translate the cost function to a Hamiltonian form. Because $\{M_i\}$ are measured quantities, they are constants, so that minimizing (\ref{eqC}) is equivalent to minimizing
\begin{equation}
    H_c=\sum_i \frac{3}{2} N_i^2-N_i\left(N_{i+1}+N_{i-1}-\frac{N_{i+2}}{2}\right)-N_i D_i
\end{equation}
with $D_i=3 M_i-2 M_{i+1}-2M_{i-1}+M_{i+2}/2+M_{i-2}/2$. Clearly, this maps to the Hamiltonian (\ref{eqH}) for Dirac 3 with
\begin{eqnarray}
    & C_i=-D_i, \\
    & J_{ii}=3/2, \\
    & J_{i,i-1}=J_{i,i+1}=-1, \\
        & J_{i,i-2}=J_{i,i+2}=1/4, \\
\end{eqnarray}
and other coefficients being zero. 

Dirac 3 currently supports 5000 modes, each holding up to 100 effective photons. This allows exact processing of LiDAR images with up to 5000 pixels. To process larger images, one may divide the image into smaller blocks, and apply the noise reversal individually in each block. This method neglects the spatial correlation between blocks and may give sub-optimum solutions. A way of mitigation is to iterate the process multiple times, each with a new set of shifted blocks.      

\section{Results}
\begin{figure}[htbp]
    \centering
    \includegraphics[width=0.95\textwidth]{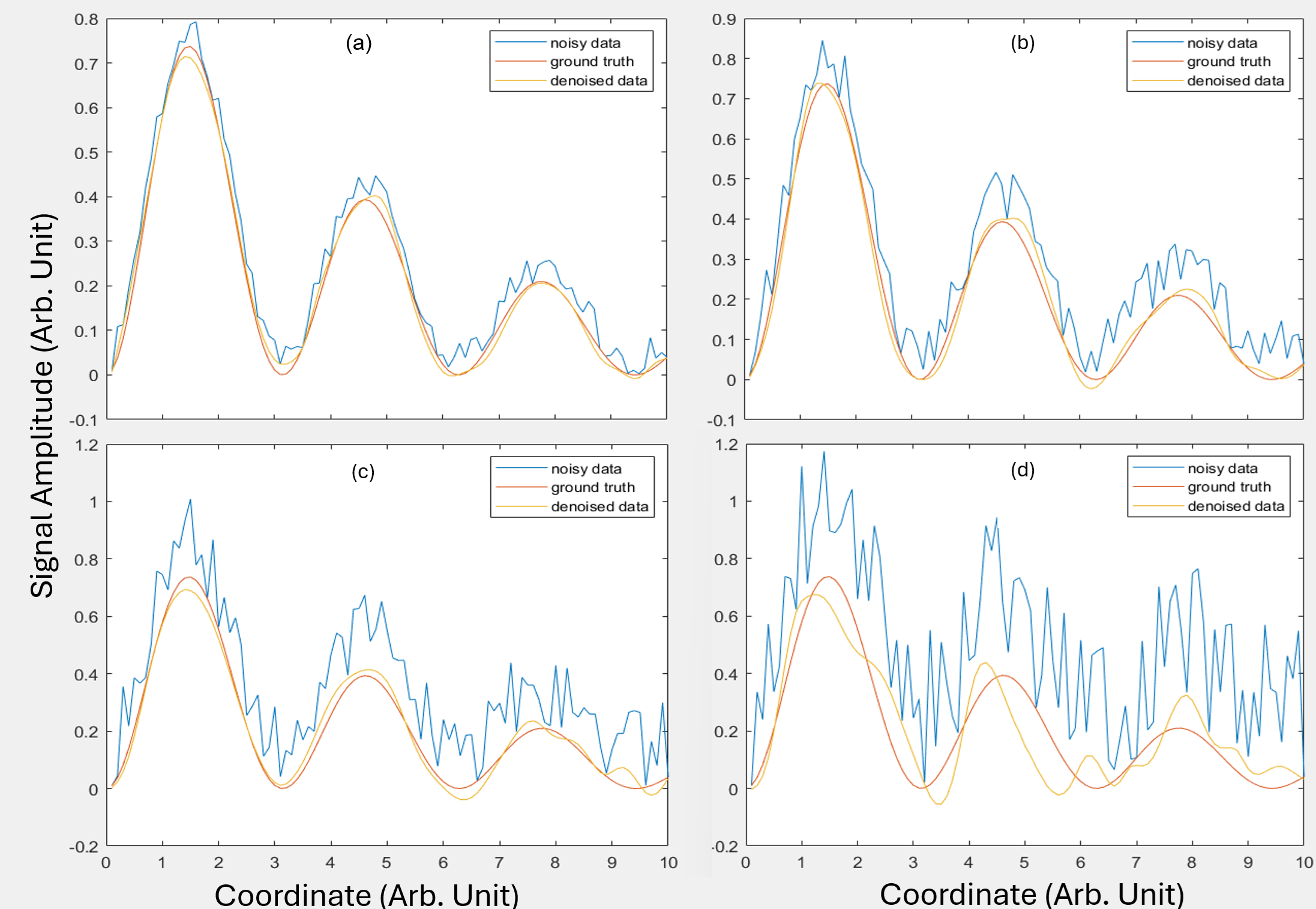} 
    \caption{Noise reversal examples of 1D data. (a), (b), (c), (d) show the ground truth, noisy data, and the noise reversal results when the noise mean amplitude is 10\%, 20\%, 40\%, and 80\% of the peak pixel value in the ground truth, respectively.}
    \label{Fig.result1}
\end{figure}

To test the noise reversal with EQC, we first consider a 1D case, where the ground truth is a decaying sinusoid, and noise of different strength is added to it for construct measured noisy data. The results, obtained by simulating the mean-field process in Dirac 3, are shown in Fig.~\ref{Fig.result1}. As seen, at 10\% and 20\% noise level, the ground truth is nearly perfectly recovered. As the noise increases to 50\%, the ground truth is still well recovered, albeit with some small deviations and a fictitious peak added at the tail, where the noise amplitude is much stronger than the signal. When the noise is at the 80\% level of the peak signal (the noise energy is larger than the signal noise), the signal is still partially recovered, with the main main peak features preserved. This results clearly demonstrate the effectiveness of EQC for noise reversal. 

\begin{figure}[htbp]
    \centering
    \includegraphics[width=0.95\textwidth]{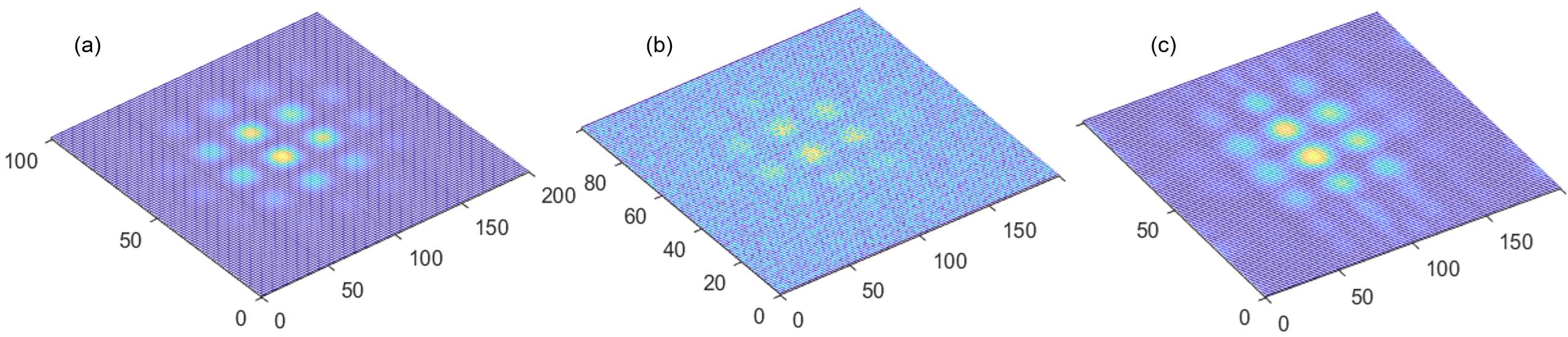} 
    \caption{Noise reversal examples of 2D data, where (a) is the ground truth, (b) plots noise-added data where the average noise amplitude is 50\% of the peak signal value in the ground truth, and (c) is the recovered data after noise reversal.}
    \label{Fig.result2}
\end{figure}

Next we test on 2D data. As shown in Fig.~\ref{Fig.result2}, the ground truth is a two-dimensional decaying sinusoid, in dimensions of 100-by-200 pixels. After significant noise is added, the image becomes noisy, with only main peaks recognizable. After noise reversal, however, the image is well restore with all peaks clearly visible. Note that there are some artificial feature appearing along the periphery. This is because there the noise is much stronger than the signal, making it impossible to recover the ground truth. 

\begin{figure}[htbp]
    \centering
    \includegraphics[width=0.95\textwidth]{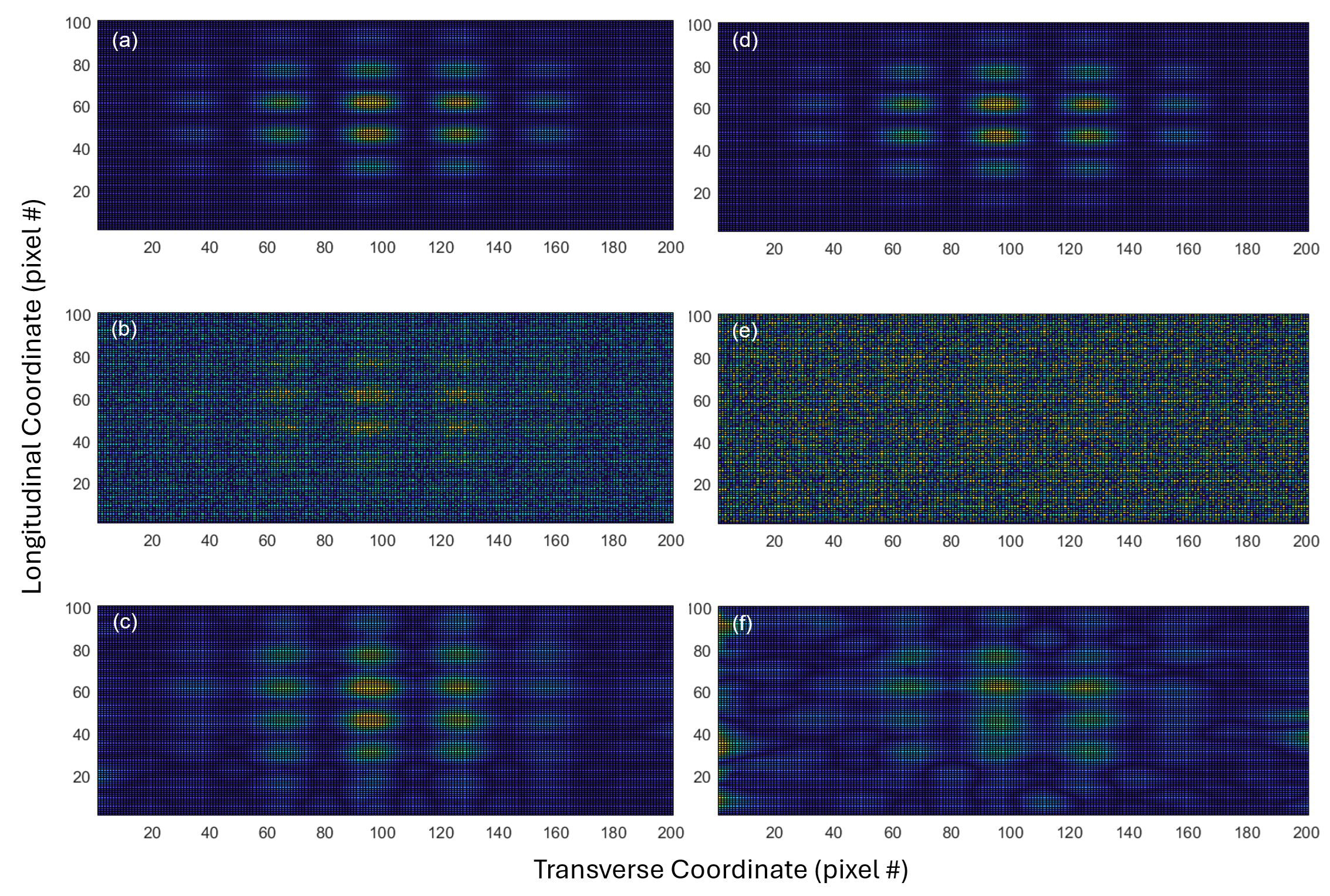} 
    \caption{Noise reversal examples of 2D data, with (a)-(c) showing the ground-truth signal, the signal mixed with noise of average amplitude equal to the peak signal amplitude, and the recovered signal after noise reversal, respectively. (e)-(c) show the same but with the average noise amplitude twice of the peak signal amplitude.}
    \label{Fig.result3}
\end{figure}

Figure~\ref{Fig.result3} plots more results for the same ground truth but mixed with stronger noise. In Fig.~\ref{Fig.result3} (a-c), the noise' average amplitude equals the signal's peak amplitude. As seen, such noise masks the signal so well that only the center lopes remain visible but all side lopes are nearly vanished. The noise reversal works quite well to almost perfectly restore the signal, with all features recovered. In Fig.~\ref{Fig.result3}(d-f), the noise' average amplitude is twice the peak amplitude. As a result, the noisy image becomes so fuzzy that even the main lopes are not visible. Yet, the noise reversal is still able recover the main features, including all the main lopes in the center. These results demonstrate the effectiveness of our approach. Note that, just like the 1D case above, some artificial lopes appear, because the noise is much stronger than the signal in those regimes. 

Comparing the 1D results in Fig.~\ref{Fig.result1} and 2D Fig.~\ref{Fig.result3}, the noise reversal performance is significantly higher under similar noise levels. This is because in the 2D case, there is correlation not only along one dimension (one column), but also across columns. EQC uses this additional correlation to better retrieve the noise and recover the ground truth.  

\begin{figure}[htbp]
    \centering
    \includegraphics[width=0.95\textwidth]{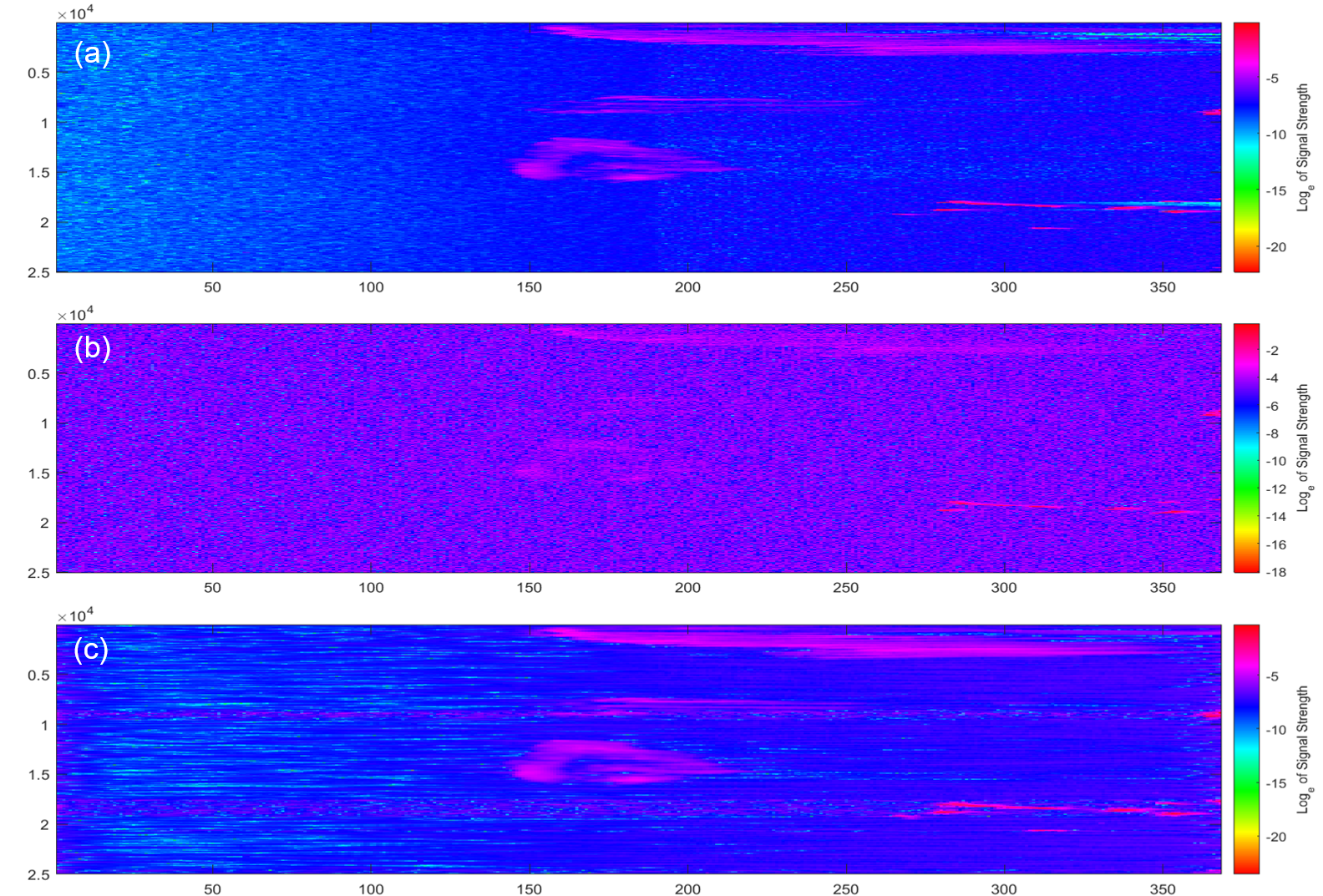} 
    \caption{Noise reversal examples of a space-LiDAR image, with (a), (b), and (c) showing the ground-truth signal, the signal mixed with noise of average amplitude is 20 times higher than that of the signal, and the recovered signal after noise reversal, respectively.}
    \label{Fig.result4}
\end{figure}

Finally, we present some preliminary results for realistic space-LiDAR data, where a photon-counting image taken at night is used as the ground truth; see Fig.~\ref{Fig.result4}(a). To simulate the daytime measurement, Poissionian random noise is added to each pixel, with the average amplitude about 20 times higher than that of the signal. The resultant image is shown in Fig.~\ref{Fig.result4}(b), where most features are lost. Note that in this case, because the signal has sharp peaks, the maximum noise is actually less than the peak value of the signal. Then, noise reversal is applied to find the noise photon number in each pixel and subtract it from the mixed image. Note here, because the image size is too large, the noise reversal is applied column by column. The spatial correlation among columns is accounted for iteratively, with a variance term added to the 1D cost function for each pixel that is the square of the difference tween the pixel and the average of the left and right pixels. The recovered image is shown in Fig.~\ref{Fig.result4}(c), where all main features are precisely restored. Again, there are small artificial features created in the low-signal area, where the noise is much stronger than the signal. 

Lastly, we try the extreme case, where the noise is 200 times stronger than the signal. This results in a completely masked image, with no feature recognizable any more. Surprisingly, EQC is still able to recover the main features. The problem, however, is that there are lots of artificial features at the left and right ends. To mitigate such errors is a main task of our future studies. 

\begin{figure}[htbp]
    \centering
    \includegraphics[width=0.95\textwidth]{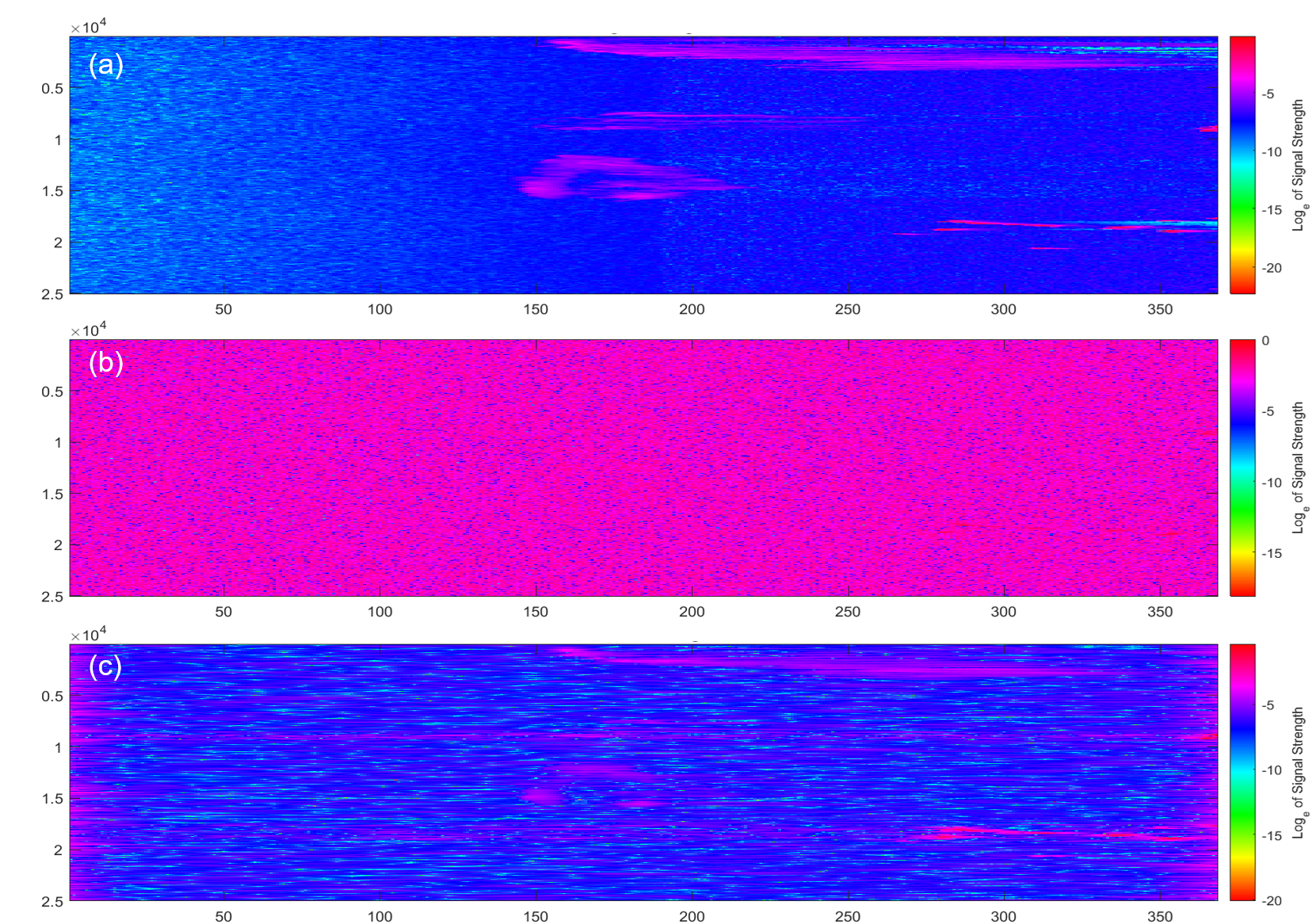} 
    \caption{Noise reversal examples of a space-LiDAR image, with (a), (b), and (c) showing the ground-truth signal, the signal mixed with noise of average amplitude is 200 times higher than that of the signal, and the recovered signal after noise reversal, respectively. }
    \label{Fig.result5}
\end{figure}

\section{Conclusion}
In conclusion, we have proposed and examined a hardware approach to noise removal using entropy quantum computing. Unlike any existing denoising method, it uses a quantum emulator to retrieve the noise according to their quantum statistical properties different from that of the true signals. As such, it can retrieve the noise and recover the signal even when it is mixed with noise that can be much stronger. This technique has broad applications in image processing, including those of LiDAR, Radar, and infrared imaging. Applying to space LiDAR missions, it is expected to significantly reduce the SWaP-C parameters of the satellites by circulating the need for high power laser. 

The results presented in this paper are just preliminary, utilizing only the mean-field effects in EQC. Much more effort is in order to quickly advance its technology readiness level for field and space missions. The first future study is to handle large-size images with mega pixels or more. This would require either EQC machines with millions of time bins or developing an effective ``stitching'' method to process an image in blocks. The second is reduce the errors in the image edges. The third is to further reduce the SWaP-C of the machines and possibly re-design them for space qualification.

\subsection* {Disclosures}
YPH is affiliated with Quantum Computing Inc. 

\subsection* {Code and Data Availability}
Data supporting the findings of the article are not publicly available at this time but can be obtained
from the authors upon reasonable request.

\subsection* {Acknowledgments}
We thank Carl Weimer from the BAE systems, Xunbin Zeng from the University of Arizona, and Charles Gatebe from NASA AMES Center for inspirational discussions, and thank Lac Nguyen, Irwin Huang, and Ting Bu at Quantum Computing Inc. for discussions and help in data processing.


\bibliography{ref}

\end{document}